\documentstyle[prb,aps,twocolumn,epsf]{revtex}
\begin{document}
\draft 
\title{Electron correlation induced transverse delocalization and 
longitudinal confinement in excited states of phenyl-substituted 
polyacetylenes}

\twocolumn[
\hsize\textwidth\columnwidth\hsize\csname@twocolumnfalse\endcsname

\author{Haranath Ghosh$^{1}$, Alok Shukla$^{1,2}$ \cite{nadd} and 
Sumit Mazumdar$^{1}$}
\address{$^1$Department of Physics and The Optical Sciences Center, 
University of Arizona, Tucson, AZ 85721}
\address{$^2$Cooperative Excitation Project ERATO, Japan Science and
Technology Corporation (JST)} 

\maketitle

\begin{abstract}
Electron-electron interactions in general lead to both ground state and
excited state confinement. We show, however, that in phenyl-substituted 
polyacetylenes electron-electron interactions cause enhanced delocalization
of quasiparticles in the optically excited state from the backbone polyene 
chain into the
phenyl groups, which in turn leads to enhanced confinement in the chain
direction. This co-operative delocalization--confinement lowers the energy
of the one-photon state and raises the relative energy of the lowest
two-photon state. The two-photon state is slightly below the optical state
in mono-phenyl substituted polyacetylenes, but above the optical state in
di-phenyl substituted polyacetylenes, thereby explaining the strong
photoluminescence of the latter class of materials. We present a detailed 
mechanism of the crossover in the energies of the one- and two-photon states
in these systems.
In addition, we calculate the optical absorption spectra over a wide
wavelength region, and make specific predictions for the polarizations of 
low and
high energy transitions that can be tested on oriented samples.
Within existing theories of light emission from $\pi$-conjugated polymers,
strong photoluminescence should be restricted to materials whose optical
gaps are larger than that of trans-polyacetylene. The present work show that
conceptually at least, it is possible to have light emission from systems
with smaller optical gaps.
\end{abstract}
]

\section{Introduction}

The absence of photoluminescence (PL) in linear polyenes, 
trans-polyacetylene (t-PA)
and the polydiacetylenes (PDAs) is a well understood electron correlation
effect. Excited states in these centrosymmetric systems have A$_g$ or
B$_u$ symmetry, which can be further classified according to their 
charge-conjugation symmetry \cite{kohler_review}. 
The dipole selection rule allows
optical transitions only between A$_g$ and B$_u$ states of opposite
charge-conjugation symmetries, and the optical transition from the 1A$_g$
ground state to the 1B$_u$ state with opposite charge-conjugation
symmetry is strongly allowed. Due to the
moderate strength of the on-site electron-electron (e-e) interaction
(the Hubbard $U$) in these
systems, however, the lowest dipole-forbidden 2A$_g$ state occurs below 
\cite{kohler_review,Ohmine,Ramasesha,Tavan} the 1B$_u$, and the optically pumped
1B$_u$ decays in ultrafast times to the 2A$_g$, radiative transition from
which to the 1A$_g$ is forbidden.

Strong PL from $\pi$--conjugated polymers like 
poly(para-phenylene) (PPP), poly(para-phenylenevinylene) (PPV) and their
derivatives is also by now well understood. As shown by Soos and collaborators
\cite{Soos1,Soos2} the lowest excited states of these systems can be
understood within {\it effective} linear chain models with large bond
alternation. Schematically, materials like PPP and PPV can be thought of
as obtained by replacement of all or alternate double bonds in t-PA with
benzene moieties, a process we term as ``bond substitution'', that in effect
increases the bond alternation of the effective linear chain \cite{note}.
The enhanced bond alternation leads to a reversed energy ordering
E(2A$_g$) $>$ E(1B$_u$) (where E(...) is the energy of the state in question),
which is conducive to strong
light emission.

\begin{figure}
\epsfxsize=2.0truein
\epsfysize=1.25truein
\centerline{\epsffile{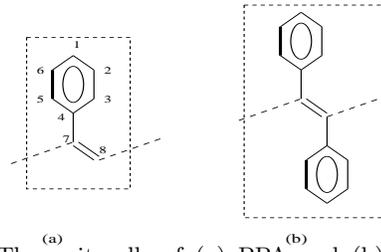}} 
\caption{The unit cells of (a) PPA and
(b) PDPA. The phenyl rings
are rotated with respect to the $y$-axis, which is transverse to the axis
of the polyene backbone ($x$-axis)}
\label{fig-unitcell}
\end{figure}

The recent observation of strong PL in phenyl- disubstituted polyacetylenes 
(PDPAs) (see Fig. \ref{fig-unitcell}) 
\cite{Tada1,Tada2,Fujii,Gontia,Sun,Hidayat}
is difficult to explain within the above simple theoretical picture. These
systems consist of a backbone polyene chain, the hydrogen atoms of which
have been replaced with phenyl groups \cite{note1}, a process that we term
as ``site substitution'', in order to distinguish it from bond substitution. 
Since the Hubbard $U$ for the 
backbone polyene carbon atoms (the repulsion between two $\pi$-electrons 
occupying the same
backbone atom), 
is an atomic
property,
it should have the same magnitude in the PDPAs and polyenes.
Since the one-electron hopping integrals between the
backbone atoms are also expected to have similar magnitudes, the observed PL
is perplexing. Several different explanations of this apparently peculiar
behavior are possible, and one goal of the present paper is to arrive at the
correct mechanism of the PL. A brief presentation of our work has been reported
earlier \cite{Shukla}. Here we present more detailed work with a broader
goal, viz., to understand the 2A$_g$ -- 1B$_u$ crossover 
that we show occurs
in these systems from a global perspective for the family of
$\pi$--conjugated polymers. Specifically, we are interested in the 
{\it mechanism} of the crossover, and not just the result. 
We show that there are two different
manifestations of the so-called
``electron correlation effect'', and as far as the relative location of the 
2A$_g$ is concerned, the two effects are competing. The effect of the
spin-dependent Hubbard correlation is to always lower the energy of the 2A$_g$
and to increase the energy of the 1B$_u$.
A second correlation effect can be the confinement of electrons and holes
in the excited 1B$_u$ state,
which has been discussed only in the context of excitons 
\cite{exciton1,exciton2}.
Although the Hubbard $U$ causes ground state localization, it has a weak
effect on the electron-hole confinement in the excited state, and the origin
of any such confinement therefore has to be different.
There can be multiple origins of this confinement, but in perhaps all these
cases the consequence is either the lowering of E(1B$_u$), or the raising of
E(2A$_g$), or both, i.e., the opposite
of the effect due to the Hubbard $U$. Our aim in the present paper is to
give a clear understanding of how this confinement occurs in the PDPAs.

A second goal of the present work is to understand and predict the
optical absorption in phenyl-substituted acetylenes over a wide wavelength
region. In the past, theoretical and experimental work on the polarizations
of the high energy absorptions in oriented samples of PPV have 
considerably increased
our understanding of electron correlation effects in this system 
\cite{Rice,Chandross,Comoretto,Miller}, and it is our hope that similar
polarized absorption measurements will be carried out on oriented samples of
phenyl acetylenes as they become available.

In order to discuss the applications of our results to real materials, we
need to clarify several aspects of the experiments with the phenyl-substituted
polyacetylenes. Before
proceeding further, we therefore present a brief overview of the experimental
situation \cite{Tada1,Tada2,Sun,Hidayat}. 
Three different classes of phenyl acetylenes have been synthesized,
which include, (i) poly-phenylacetylenes (PPAs), which are obtained by
substituting alternate hydrogen atoms of t-PA with phenyl 
groups \cite{note1}, (ii)
poly-phenylalkylpolyacetylenes (PAPAs), 
in which alternate hydrogens of t-PA
are replaced with alkyl and phenyl 
groups \cite{note1}, respectively, and (iii) the PDPAs,
already mentioned above. In Figs.  \ref{fig-unitcell}(a) and 
(b) we have shown the structures
of PPA and PDPA, respectively.
The PPAs are weak emitters, while whether
or not the PAPAs emit light depends on the size of the alkyl group, -- for 
small alkyl groups the PL quantum efficiency is small while for 
large alkyl groups this can be as large as in PDPAs (see Fig.~6 in
reference \onlinecite{Tada2}).
In contrast, all PDPAs are strong emitters of light, in the green or
blue region of the spectrum. Raman scattering
studies that have probed the backbone $C=C$ bonds have determined that the
conjugation lengths of PAPAs and PDPAs are short \cite{Fujii} (5 -- 7 double
bonds along the backbone according to the authors).
The finite conjugation length is perhaps to be anticipated, based on the
large sidegroups that are apt to lead to twists and bends involving
the backbone main chain. 
Finally, in
both PPAs and PDPAs (see Fig. \ref{fig-unitcell}) steric effects involving the 
phenyl groups
would be considerable, and moderate phenyl group rotations are to
be expected. There is currently no experimental information on the extent
of this ring torsion.

In principle, there can be multiple reasons for the observed light emission
in PDPAs. First, the light emission can be a simple consequence of the loss
of C$_{2h}$ symmetry due to ring torsions. Within this scenario 
the nominally 2A$_g$ occurs below the 1B$_u$, but in the absence of strict
symmetry classification, this state is emissive \cite{Fujii}.
Second, the energy ordering {\it is} reversed, i.e., E(2A$_g$) $>$ E(1B$_u$),
but this is a simple
consequence of finite conjugation lengths, -- it is, for example, believed that
in the shortest polyenes the 2A$_g$ might occur above the 1B$_u$. 
Third, it is conceivable that the light emission is a 
consequence of an intricate electronic mechanism that involves the electron
hopping among the phenyl groups. Referring to Fig. \ref{fig-unitcell}, for example, even with
ring torsions the distances between consecutive phenyl groups 
on the same side of the backbone polyene chain in both PPA and PDPA
are rather
small. If these are now nearly coplanar (i.e., if the ring torsions are
ordered) there should be considerable overlaps between the $\pi$-electron
clouds of the phenyl groups, leading to electron hoppings between them
\cite{Rice1}, 
as is well-known in the literature on segregated stack organic
charge-transfer solids
\cite{CTsolids}. Fourth, it is conceivable that
phenyl-substitution leads to enhanced bond alternation on the backbone polyene
chain, and this is what leads to the reversed energy ordering and light
emission, exactly as in PPV and PPP \cite{note2}. Finally, light emission
can be a consequence of confinement or localization along the backbone, which
is accompanied by enhanced delocalization into the phenyl sidegroups
\cite{Shukla}. We have ignored in the above the possibility that excited
state ordering E(2A$_g$) $>$ E(1B$_u$) is due to the relaxation of the 1B$_u$
exciton, as the 1B$_u$ -- 2A$_g$ energy gap in t-PA is known to be fairly 
large, and even if exciton relaxation plays a role, 
this would still require that this
energy gap in PDPAs has shrunk due to an electronic mechanism.

We point out that the experimental results already preclude most of the
possible scenarios. The loss of C$_{2h}$ symmetry is stronger in PPAs 
than in PDPAs, and if the light emission was from the nominally 
2A$_g$ state, one
would have expected these to be more strongly emissive than the PDPAs. 
Furthermore, in our explicit calculations for PPA, 
we have found that the transition
dipole coupling between the nominally 2A$_g$ and the nominally 1A$_g$ {\it is}
nonzero,
but its magnitude remains very small (we therefore continue to refer to
eigenstates as A$_g$ and B$_u$, even though this classification is not strictly
valid). Lack of C$_{2h}$ symmetry is therefore not the origin of the strong PL. 
Finite conjugation length alone cannot
explain the light emission, as the estimated conjugation lengths 
\cite{Fujii} are in the
region where the 2A$_g$ occurs below the 1B$_u$ in the polyenes \cite{Kohler2}.
The direct electron hopping between the phenyl groups do introduce a novel
conjugation channel absent in other $\pi$--conjugated polymers, but this
effect should be comparable in the PPAs, PAPAs and PDPAs. The absence of
strong PL in the PPAs and many PAPAs then again
precludes this mechanism. Indeed, as we show from
explicit calculations, this particular interaction affects the lowest energy
allowed and forbidden optical gaps very weakly, although it can have a 
strong effect on the optical absorption in the high energy region. Thus from
the experiments alone, we believe that there are only two possible
explanations of the strong PL in PDPAs, viz., enhanced bond alternation
and an electron correlation induced confinement-delocalization
effect. Our previous work
had already shown that the bond alternation is smaller in the PDPAs than in
t-PA. In the present work we show further detailed work, emphasizing on the
mechanism behind the reduced bond alternation, rather than the result itself.
Our aim is to show that this effect and the confinement effect are ultimately
related, and even though the 2A$_g$ -- 1B$_u$ crossover is a many-body 
problem, the final results could have been perhaps anticipated from a
combination of our understanding of one-electron theory and confinement
effects due to e-e interactions.
 
In the rest of the paper we present detailed calculations for PPA and PDPA
oligomers and polymers. We do not present explicit calculations for PAPA,
the understanding being that the effect of the alkyl groups on the electronic
structure is by and large the same as that of the hydrogen atoms, with the
only difference that the site energies (electronegativities) 
of the backbone carbon atoms to which
the alkyl groups are bonded might be weakly different. We postpone comments
on the strong PL from PAPAs with very large alkyl groups until the last
section of the paper.
In section II we present our Hamiltonian, along with a detailed discussion
of the parameters.
In section III we give a brief discussion of known results concerning the
effect of confinement on the 2A$_g$ -- 1B$_u$ crossover in linear chains. 
These discussions point out already the necessity to have a precise 
understanding of the electronic structures of PPA and PDPA within the
one-electron limit of our Hamiltonian, our results for which are presented
in section IV.
In section V, we present details of the optical absorption within the
correlated Hamiltonian within the singles-configuration interaction (SCI)
approach. The goal of this section is two-fold. First, we show that the
electron-hole confinement along the backbone has a synergistic relationship with
the delocalization into the phenyl rings.
A second goal is to make specific
predictions that can be tested out in absorption studies of oriented samples
in the future. In section VI, we
present the results of multiple reference singles and doubles
configuration interaction (MRSDCI) studies \cite{Tavan}
on short PPA and PDPA oligomers
to prove the reversed energy ordering in these systems. These calculations are
considerably more sophisticated than our previously reported calculations
\cite{Shukla}, but nevertheless, slightly longer oligomers could be investigated
here. In section VII we present our concluding discussions, 
emphasizing the fact that the mechanism of 2A$_g$ -- 1B$_u$ crossover found
here introduces the novel possibility of synthesizing light emitting 
$\pi$--conjugated polymers with small optical gaps \cite{Shukla}.

\section{Theoretical Model}
\label{theory}
Our calculations are within the $\pi$-electron model for PPAs and PDPAs,
as the ordering of the lowest energy states depends primarily on the
electron correlation effects among the $\pi$-electrons.
Even with this restriction, the complete theoretical model for PPAs and PDPAs 
is rather complicated, as this must include the effects of the 
electron-phonon (e-ph) couplings along the 
backbone polyene chain, electron-electron (e-e) interactions among all the
$\pi$-electrons, and all effects of phenyl-group rotations. Phenyl-group
rotations modify the one-electron hoppings between the phenyl groups and the
backbone chain and the intersite Coulomb interactions involving the carbon 
atoms of the phenyl groups, and in addition, introduce $\sigma$-type
bondings between the $\pi$-molecular orbitals (MOs) of neighboring phenyl 
groups on the same side of the polyene chain.
This last interaction is maximized when the rotational angle is ordered and
90$^o$, but such large rotation would completely destroy 
the phenyl-polyene conjugation,
leading to an effectively isolated polyene chain and isolated
benzene molecules. Since
the properties of PPAs and PDPAs are different from those of isolated polyene
chains, the rotation angle must be considerably less than 90$^o$, in which
case the one-electron hoppings between neighboring phenyl groups can involve
only a few of the carbon atoms of each phenyl group (see Fig. \ref{fig-unitcell}).

From the above, the overall Hamiltonian for PPAs and PDPAs is written as,
\begin{mathletters}
\label{allequations}
\begin{equation}
H = H_C + H_B + H_{CB} + H_{BB} + H_{ee},  \label{eq-ham}
\end{equation}
Here $H_C$ and $H_B$ are the one-electron Hamiltonians for the backbone
chain carbons and the benzene units, respectively, H$_{CB}$ and H$_{BB}$
are the one-electron hoppings between the chain and the phenyl
units, and between the phenyl groups themselves,
respectively, and H$_{ee}$ is the e-e
interaction. The individual terms can now be written as, 
\begin{equation}
H_C = -\sum_{\langle k,k' \rangle,M} (t_0 - \alpha \Delta_{k,M}) 
B_{k,k';M,M+1}
 + \frac{1}{2}  K  \sum_{k,M} \Delta_{k,M}^2 \label{eq-h1}
\end{equation}
\begin{equation}
H_B=-t_0 \sum_{\langle\mu,\nu\rangle,M} B_{\mu,\nu;M,M} \label{eq-h2}
\end{equation}
\begin{equation}
H_{CB}= -t_{\perp} \sum_{\langle k,\mu \rangle,M} B_{k,\mu;M,M,} \label{eq-h3}
\end{equation}
\begin{equation}
H_{BB}= -\displaystyle\sum_{\mu,\nu,M,M'} \nolimits ' t_{\sigma}(\mu,\nu) B_{\mu,\nu;M,M'} \label{eq-h4}
\end{equation}
\begin{eqnarray}
H_{ee}&=&U \sum_{i,M} n_{i,M,\uparrow} n_{i,M,\downarrow} \nonumber \\
& &
+ \frac{1}{2} \sum_{i \neq j,M,N} V_{i,j,M,N}(n_{i,M}-1)(n_{j,N}-1) \label{eq-hee}
\end{eqnarray}
\end{mathletters}
In the above, $k$, $k'$ are carbon atoms on the polyene backbone, 
$\mu, \nu$ are carbon atoms belonging to the phenyl groups, $M$ is a
composite site 
consisting of a phenyl group and a polyene carbon, $\langle ... \rangle$
implies nearest neighbors, and 
$B_{i,j;M,M'} = \sum_{\sigma}(c_{i,M,\sigma}^\dagger c_{j,M',\sigma} + 
h.c.)$. The $t_0$, $t_{\perp}$ and $t_{\sigma}(\mu, \nu)$ are matrix elements 
corresponding to one-electron hops, where $t_{\sigma}$ depends on the
particular $\mu$, $\nu$ being considered.  
In $H_C$,  
$\alpha$ is the e-ph
coupling constant, $K$ the spring constant, and 
$\Delta_{k,M}=(u_{k+1,M+1}-u_{k,M})$, where $u_{k,M}$ is the displacement
of the Mth composite site from equilibrium. 
In $H_{CB}$, the sum over $\mu$ is restricted
to atoms of the phenyl groups that are directly bonded to backbone carbon
atoms. The prime
on the summation in $H_{BB}$ indicates that the sum is restricted to a few
carbon atoms of neighboring phenyl groups that are on the same side of the
polyene chain ({\it i.e.}, $M' = M \pm 2$), the actual number depending upon
the extent of phenyl group rotation.
$H_{ee}$ is the e-e interaction, with $i$ and $M$ including now all
atoms. The Coulomb interactions are parametrized according to  
the Ohno relationship \cite{ohno},
\begin{equation}
V_{i,j,M,N} = U/(1+0.6117R_{i,j,M,N}^2)^{1/2} \; \mbox{,}
\label{eq-ohno}
\end{equation}
where $U$ = 11.13 eV and $R_{i,j,M,N}$ is the distance in \AA ~ between the
$i$th carbon in unit $M$ and the $j$th carbon in unit $N$.
The Ohno parametrization is for small molecules, and it is conceivable that
the Coulomb parameters for the polymeric samples are somewhat smaller 
\cite{Chandross1}. 
Our aim, however, is to obtain a {\em qualitative} understanding
of the role played by e-e interactions on the optical
properties of the system. We have therefore employed large $H_{ee}$ to
show that in spite of these large interactions energy ordering
E(2A$_g$) $>$ E(1B$_u$) will occur in the substituted
polyacetylenes. The same considerations apply to the other parameters 
(see below), viz., 
the actual numerical values are relatively unimportant, as our goal is
to extract the essential physics of these systems. 

In our calculations the backbone polyene 
atoms
were assumed to lie 
in the $xy$ plane, with the $x$ direction defined as the 
polymer axis (see Fig.~1). 
Due to
steric effects, the phenyl rings will not be in the $xy$ plane,
but will be rotated with respect to the $y$ axis.  
All intra-phenyl bond 
lengths are taken to be 1.4 \AA ~ and the bond between the backbone atoms 
and the phenyl groups is taken to be a
true single bond with length 1.54 \AA. In our calculations of bond-alternation
along the polyene backbone (section IV), 
the backbone polyene bond lengths are assumed to be
1.4 \AA $~$in the undistorted configuration, while the bond lengths 
are calculated self-consistently in the distorted configuration.
The calculations of optical absorption are for finite oligomers, in which
we chose the backbone bond lengths the same as in t-PA, viz.,
1.45 \AA $~$ and 1.35 \AA, for reasons to be discussed later. 
We considered both parallel and
antiparallel ordered rotations of the phenyl groups bonded to the nearest
neighbor carbon atoms in PDPA (note that these occur on opposite sides of
the backbone, the phenyl groups on the carbon atoms on the same side of the
chain are parallel in both cases)
with rotation angles of 30$^o$ (this value of the torsional angle is of
course arbitrary; on the other hand torsional angles larger than 45$^o$
will decrease $t_{\perp}$ drastically). However, the differences in 
numerical results (energies as well as transition dipole couplings)
for the two cases were insignificant. Therefore, we restricted our studies
to the parallel configuration for the phenyl groups, 
rendering the symmetry group of PDPA to be $C_i$, with identity and inversion
operators as symmetry elements.
PPA, on the other hand, does not possess any point group 
symmetry. 

As far as the hopping matrix elements are concerned, we assumed 
$t_0$ = 2.4 eV.
Along the backbone carbon atoms, whenever rigid bond alternation was
assumed along the chain, standard values of 2.2 eV and 2.6 eV were chosen
for the long
and the short bonds, respectively.
The aforesaid rotation of the phenyl groups with respect to the $y$ axis 
will have the effect of reducing the hopping integral $t_{\perp}$
connecting the phenyl rings to the t-PA backbone, as compared to the standard
value of 2.4 eV. Various possible values for $t_{\perp}$ were considered
in our calculations. We postpone the discussion of our choice of
$t_{\sigma}(\mu,\nu)$ until later.

\section{Confinement and the 2A$_{\lowercase{g}}$--1B$_{\lowercase{u}}$ crossover}

Although the theory of 2A$_g$ -- 1B$_u$ crossover is well understood in linear
chains, it is useful to present a brief review in view of what follows. 
Consider the rigid bond limit of Eq.~(1) for the case of linear chains,
with $t_{\perp}$ = 0 and
the chain hopping integrals $t_0(1 \pm \delta)$.
In the limit of large $U$, the ground state has all atoms singly occupied
by electrons and the 2A$_g$ is a spin excitation that is sum of
two triplets, each localized on the double bonds, and has the energy
4$t^2(1 + \delta)^2/(U - V)$, where $V$ is the Coulomb interaction 
between nearest
neighbor atoms on the linear chain \cite{kohler_review,Ohmine,Ramasesha,Tavan}. 

For finite $U$, the overall wavefunction
is a superposition of this spin excitation and a charge excitation, and
the 2A$_g$ occurs below the 1B$_u$ if the contribution of the spin excitation
configuration dominates
over that of the charge excitation 
configuration. From configuration interaction principles, the
relative contribution of the spin excitation decreases as its energy
increases, and
therefore both increasing $\delta$ (for fixed $U$ and $V$) and increasing
$V$ (for fixed $U$ and $\delta$) will increase the energy of the 2A$_g$
relative to the 1B$_u$. This can be understood physically in the limit of
large $\delta$ or $V$ rather easily. The spin excitation that is relevant in 
the description of the 2A$_g$ is obtained

\begin{figure}[htb]
\twocolumn[\hsize\textwidth\columnwidth\hsize\csname @twocolumnfalse\endcsname
\epsfxsize=6.25truein
\epsfysize=2.25truein
\centerline{\epsffile{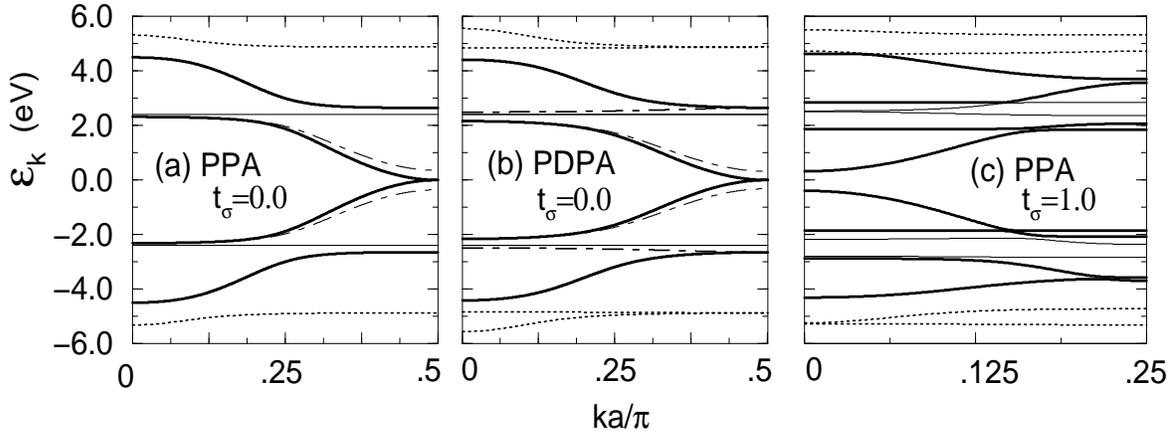}  }
\medskip
\caption{Band structures of (a) PPA and (b) PDPA for $t_{\perp}$=1.4 eV
and $t_{\sigma}$ = 0, and (c) PPA for $t_{\perp}$=1.4 eV and 
$t_{\sigma}$ = 1.0 eV. 
Here thick solid lines represent polyene chain-derived bands, with zero
bond alternation in (a) and (b), and with finite bond alternation in (c).
Finite
bond alternation in (a) and (b) opens a gap at the Fermi level, and the
highest valence and lowest conduction bands in these cases are shown as
thin dashed--dotted lines.
The thin solid lines represent benzene-derived localized bands (see text)
and the thick dashed--dotted lines in (b) represent benzene-derived 
delocalized bands. There occur
delocalized benzene-derived bands in (a) which are nearly degenerate with the
flat localized bands and are not visible within our resolution.} 
\label{fig-band}
\medskip]
\end{figure}

\noindent from the localized ground state by
two consecutive electron hops in
opposite directions, the first creating a virtual state 
with a doubly occupied site and a vacancy,
and the second destroying this. Now increasing $\delta$ or $V$ both 
have the tendency to confine the double occupancy and the vacancy in the
virtual state to nearest neighbor dimer ethylenic units, and since the dimer
does not have a spin singlet covalent excitation, there always exist
a $\delta$ or $V$ where E(2A$_g$) $>$ E(1B$_u$), with the only condition
that $\delta$ is nonzero to begin with. Thus any interaction that confines
the double occupancy and the vacancy in the virtual state should increase
the {\it relative} energy of the 2A$_g$. We refer to references 
\onlinecite{Soos1,Soos2} for further details on the effect of $\delta$, but 
point out a fundamental difference between the effect of $\delta$ and $V$
on the 1B$_u$. While increasing $\delta$ (for fixed $U$ and $V$)
increases E(1B$_u$), increasing $V$
(for fixed $U$ and $\delta$) decreases E(1B$_u$) (since the correlation
contribution to the optical gap is linear in $U - V$). Although 
increasing $V$ is not a practical proposition in linear chains, the more
important point is that in principle at least, energy ordering conducive to
light emission does not necessarily require that E(1B$_u$) be larger than
in the polyenes or t-PA.

Recognition that any interaction that promotes confinement of the 
vacancy and the double occupancy in the virtual state allows us to 
understand the {\it mechanism} of a possible 2A$_g$ -- 1B$_u$ crossover in
PDPA by probing the 1B$_u$ in detail, since the same interaction will also
lead to confinement in the 1B$_u$. Now,
probing the 1B$_u$ theoretically is much simpler than probing the
2A$_g$. This is because the SCI approximation, which is easy to implement,
is known to give a reasonable
description of the 1B$_u$, but not that of the 2A$_g$. Any possible
confinement in the
1B$_u$ in PPA and PDPA can only occur due to conjugation with the phenyl 
sidegroups, and in order to understand the effects of this
conjugation in detail we have to
start from one-electron theory, as is done in the next section. 

\section{One-electron theory}
\label{huckel}
In this section we present the results of our calculations of
band structures, bond alternations, and optical absorptions in PPA and
PDPA based upon one-electron theory ($H_{ee}$ = 0 within Eq.~(1)). 
It will be seen that the difference
between PPA and PDPA on the one hand, and polyenes on the other, can be
largely understood already within one-electron theory, and the effects of
e-e interactions can be anticipated based upon these results.

\subsection{Band Structure}
\label{band-s}
The band structures were calculated within the rigid bond approximation,
and two different cases, viz., equal bond lengths 
and bond lengths corresponding
to those of t-PA along the polyene backbone 
were considered. 
Furthermore, since the number of carbon atoms to be retained in $H_{BB}$ is
somewhat arbitrary (as the exact extent of the overlap between the $\pi$-MOs
of neighboring phenyl groups is difficult to estimate) we perform two different
sets of band structure calculations, with $H_{BB}$ both zero and nonzero.
As these calculations indicate, the strongest effects of phenyl substitution 
in the region close to the Fermi level are those due to $H_{CB}$, 
with $H_{BB}$ playing a smaller quantitative role. $H_{BB}$, however, plays
a stronger role away from the Fermi level.

For $H_{BB}$ = 0 the one-electron part of Hamiltonian ~(1) is the simple
H\"uckel model. In this case, the consequence of nonzero $H_{CB}$ is
simple hybridization between 
the polyene bands and benzene molecular orbitals (MOs). The highest (lowest)
occupied (unoccupied) benzene MOs (HOMOs and LUMOs) 
are doubly degenerate, with one member of each set of
doubly degenerate levels being delocalized over all the carbon atoms of the
benzene molecule and the other localized over four of the six carbons
\cite{Salem}. Henceforth, we refer to the delocalized HOMO (LUMO) as $d$ 
($d^*$) and the localized HOMO (LUMO) as $l$ ($l^*$). We
expect very strong hybridization between the polyene valence (conduction)
bands and the $d$ ($d^*$) MOs, but the $l$ and $l^*$ MOs are expected
to retain their characters, as in the case of PPV 
\cite{Rice,Chandross,Chandross1,Cornil}. 
In addition
to the $d$, $d^*$ frontier MOs, each benzene molecule also has a 
delocalized low (high)
energy occupied (unoccupied) MO, hybridization between which and the 
polyene-derived bands
is expected to be smaller (except at the edge of the Brillouin zone) 
due to the large one-electron energy difference.

In Figs. \ref{fig-band} (a) and (b) we have shown the band structures 
for PPA and PDPA,
respectively, for $H_{BB}$ = 0, but nonzero $H_{CB}$, with 
$t_{\perp}$ = 1.4 eV. 
We have verified that the band structure remains qualitatively the same for
all $t_{\perp}$ between 1 eV and 2.0 eV. For the case of zero bond 
alternation, there exists a Fermi surface degeneracy, exactly as in the case
of t-PA, indicating unconditional bond alternation in the infinite chain
limit. The band structures are nearly identical for zero and nonzero bond 
alternation, except at the center of the Brillouin zone, 
where a gap opens up for nonzero
bond alternation in both PPA and PDPA.

For phenyl group rotations of 30$^o$ - 45$^o$, $H_{BB}$ has
significant contributions only from the overlaps between the carbon atoms
2 and $6^\prime$, and between atoms 3 and $5^\prime$ in Fig. 
\ref{fig-unitcell} (here $6^\prime$ and $5^\prime$ refer to carbon atoms
equivalent to 6 and 5, on the next phenyl group on the same side of the
polyene chain).
We will therefore ignore
the more distant $t_{\sigma}(\mu,\nu)$ and will henceforth refer to the
nearest neighbor terms that we retain as simply $t_{\sigma}$. In the 
absence of
theoretical and experimental information concerning the magnitudes of 
these
hopping integrals, we have chosen relatively large values of 1.0 eV and
2.0 eV, to determine the largest effects of $H_{BB}$.
The band structure of PPA
for the case of $t_{\sigma}$ = 1.0 eV is shown in Fig. \ref{fig-band}(c).
There are two effects of nonzero $H_{BB}$. First, charge conjugation
symmetry is broken. Second, there is now much larger dispersion in the
``benzene-derived'' bands. However, the gap between the highest valence
band and the lowest conduction band remains largely
unaffected, a result that remains true even for $t_{\sigma}$ = 2.0 eV
(not shown).
Due to the
doubling of the unit cell upon inclusion of $H_{BB}$ the Brillouin zone 
is halved and the number of bands for PPA is twice those in 
Fig. \ref{fig-band}(a). In the
case of PDPA this number is larger by almost another factor of two, and we
have therefore not shown the band structure for PDPA for nonzero $t_{\sigma}$.
As in PPA, however,
$H_{BB}$ has an insignificant effect on the energy gaps in the low energy
region.
We therefore do not expect the low energy physics (especially
the energy ordering upon inclusion of $H_{ee}$) to be altered by $H_{BB}$.
On the other hand, because of the larger dispersion in the benzene-derived
bands, nonzero $H_{BB}$ leads to
chain-to-benzene charge-transfer absorptions at relatively lower energy
(see below).

From the band structures alone, the strong hybridization that occurs
between the polyene bands and the benzene delocalized HOMO and LUMO are
obvious only away from the center of the reduced Brillouin zone
($k = \pi/2a$ in Figs. 2(a) and (b) and $k$ = 0 in Fig.~2(c)). 
A more quantitative measure of the
hybridization, however, is the contribution of the chain atoms to the band
wavefunctions. 
We have calculated this quantity for the one-electron wavefunction
corresponding to the 
Fermi level $k_F$, defined as,  
${\rho_{c,k_F} =  \sum_{i \in chain} c^*_{k_F,i} c_{k_F,i}}$, 
where $c_{k_F,i}$ is the coefficient of the $i$th chain atom to the 
wavefunction at $k = k_F$, 
for both PPA and
PDPA as a function of $t_{\perp}$. In Fig. \ref{fig-chain-bond}(a)
we have summarized our results,
where we see that ${\rho_{c,k_F}}$ decreases
rapidly with $t_{\perp}$. Similar calculations were done also for 
one-electron levels away from $k_F$. 
In general, the chain contribution to
the highest occupied and lowest unoccupied bands 
decreases with increasing absolute value of the one-electron energy
(there of course exists a sum rule, which implies that the nominally
benzene-derived bands away from the Fermi level have considerable contribution
from the polyene chain atoms). Since the Peierls bond-alternation in t-PA
is a strictly one-dimensional effect, the large decrease in
$\rho_{c,k_F}$ with $t_{\perp}$ indicates that the bond-alternation 
in PPA and PDPA
should be smaller than in t-PA.

\subsection{Bond Alternation}
\label{sec-ba}
\begin{figure}[htb]
\twocolumn[\hsize\textwidth\columnwidth\hsize\csname @twocolumnfalse\endcsname
\epsfxsize=4.25truein
\epsfysize=2.5truein
\centerline{\epsffile{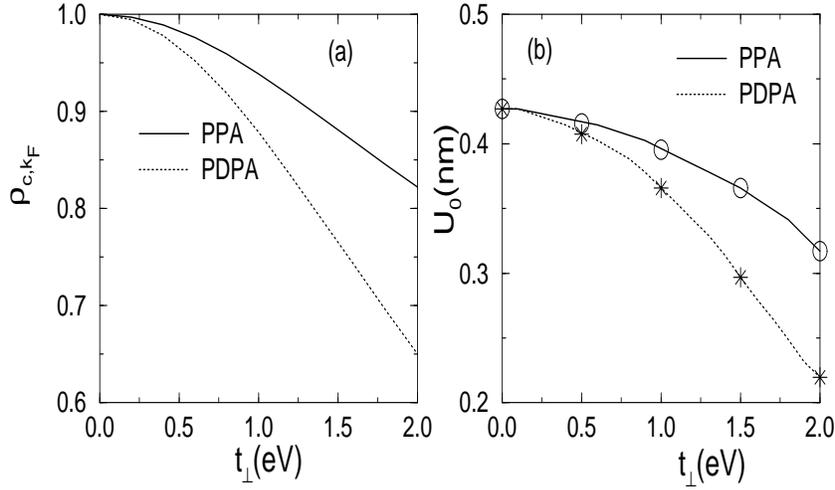}} 
\medskip
\caption{(a) The contributions by the polyene chain atoms to the one-electron
wavefunction at $k = k_F$ in PPA and PDPA, as a function of
$t_{\perp}$ for $t_{\sigma}$ = 0. Nonzero $t_{\sigma}$ has almost no
effect on $\rho_{c,k_F}$. (b) The self-consistent bond alternation 
(in nanometers) within the
one-electron limit of Eq.$~$(1), as a function of $t_{\perp}$. The continuous
curves are for $t_{\sigma}$ = 0, while the stars ($\star$) and 
circles ($\circ$)
correspond to $t_{\sigma}$
= 1.0 eV for PDPA and PPA, respectively, 
again indicating the insignificant effect of $t_{\sigma}$ on the
low energy properties.}
\label{fig-chain-bond}]
\end{figure}

For bond alternation within a periodic configuration,
$u_{k,M} = (-1)^Mu_0$ in Eq.~(1).
We have calculated the optimal bond alternation 
$u_0$, within the one-electron limit ($H_{ee}$ = 0) of Eq.~(1).
The magnitude of
$u_0$ depends on the dimensionless e-ph coupling constant $\alpha^2/Kt_0$,
but the qualitative behavior as a function of $t_{\perp}$ does not. 
We have therefore chosen the original Su-Schrieffer-Heeger \cite{SSH}
values of $\alpha$
(4.1 eV/\AA) and K (21 eV/\AA$^2$), for direct comparisons to previous
one-electron theories of t-PA. In reference \onlinecite{Shukla}, 
we had shown the behavior of
finite PDPA rings with variable size for a few fixed $t_{\perp}$. Here we
show the behavior of both PPA and PDPA for $0 < t_{\perp} < 2.0$ eV in 
Fig. \ref{fig-chain-bond}(b).
These results are for periodic 
rings containing 100 backbone carbon atoms, where the optimal
$u_0$ has reached its saturation value in all cases. The 
close relationship between
the optimal $u_0$ and $\rho_{c,k_F}$  
is obvious from the natures of these curves. 

As has been argued in reference \onlinecite{Shukla}, 
nonzero $H_{ee}$ will change
the magnitude of $u_0$ in each polymer, but will not change the trend seen
in Fig. \ref{fig-chain-bond} (b), viz., a reduction in the bond alternation 
with increasing $t_{\perp}$. We therefore conclude that the
observed PL in
PPA and PDPA cannot be a consequence of large effective bond 
alternation, as is true in the bond-substituted materials PPP and PPVs.

%

\subsection{Optical absorption}
\label{sec-og}
Although our interest lies primarily in calculating the optical absorption
with nonzero $H_{ee}$, it is useful to discuss the $H_{ee}$ = 0 limit first. 
In the
case of PPV, comparisons of absorption spectra calculated within one-electron
and many-electron theories on the one hand, and with experimental spectra on the
other \cite{Chandross,Comoretto}, have led to a clear understanding of the
effects of the e-e interaction. 
Although oriented samples of phenyl-substituted polyacetylenes do not 
exist at the moment, we give details of the calculated polarizations here,
in anticipation of future experimental
work.

The optical absorption within one-electron theory can be anticipated from
the structures of PPA and PDPA. In the absence of $H_{CB}$ and $H_{BB}$ we
expect isolated low energy polyene chain absorption polarized primarily 
along the 
$x$-axis, and degenerate
benzene transitions at high energy which can be characterized as 
$d \to d^*$ ($y$-polarized),
$l \to l^*$ ($y$-polarized) and $d \to l^*$ and $l \to d^*$ ($x$-polarized)
(these polarizations can be anticipated from previous work on PPV, as the
unit cell in PDPA is simply the trans-stilbene molecule).
The anticipated effects of nonzero $H_{CB}$ are: (i) enhanced $y$-character
in the lowest energy chain absorption due to hybridization with the benzene
delocalized MOs, (ii) splitting of the high energy molecular benzene 
transitions, and (iii) new degenerate chain $\to$ benzene and benzene $\to$
chain charge-transfer type absorptions at intermediate energies. The 
charge-transfer type absorptions will include transitions from and to
benzene $d$, $d^*$ and $l$, $l^*$ orbitals. Of these, the former 
(involving $d$, $d^*$ MOs) are
expected to have mixed polarization (since these are hybridized with the
chain-derived bands that have some $y$-character, see above), but the latter
(involving $l$, $l^*$ MOs) should be primarily
$x$-polarized. From the band structures in Figs. \ref{fig-band}(c), 
for nonzero $H_{BB}$
we expect even larger splitting in the molecular transitions, and thus the
width of the absorption band in this region should increase. Due to broken
charge-conjugation symmetry, the chain $\to$ benzene and benzene $\to$ chain
absorptions are also nondegenerate, and therefore the absorption band in the
intermediate energy should also be broader.

\begin{figure}[htb]
\twocolumn[\hsize\textwidth\columnwidth\hsize\csname @twocolumnfalse\endcsname
\epsfxsize=6.truein
\epsfysize=3.5truein
\centerline{\epsffile{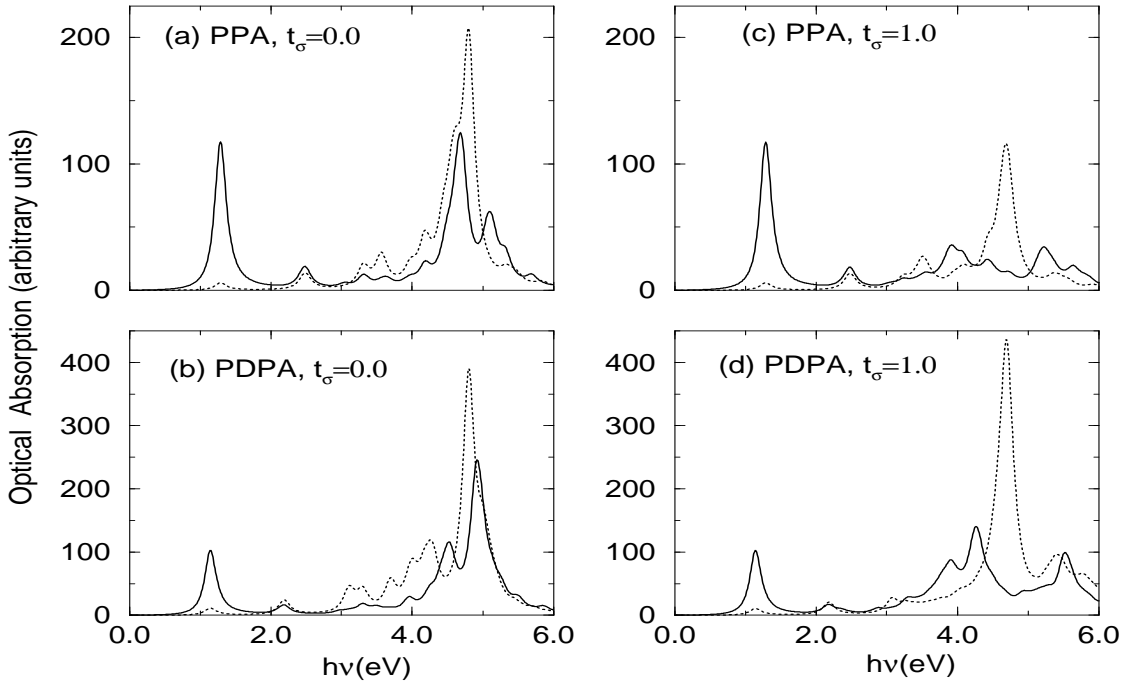}} 
\medskip
\caption{The calculated $x$ (solid line) and $y$ (dotted lines) components
of the optical absorption spectra
within one-electron theory for finite PPA (Figs.~(a) and (c)) and PDPA
(Figs.~(b) and (d)) oligomers with eight double bonds on the backbone polyene
chain. The y-axis scales are different for PPA and PDPA.
In all cases $t_{\perp}$ = 1.4 eV, while $t_{\sigma}$ = 0 in Figs.
(a) and (b) and 1.0 eV in Figs.~(c) and (d)).}
\label{huk-alphaxy-tb}
\medskip]
\end{figure}

In Figs. \ref{huk-alphaxy-tb} (a) and (b) we have shown 
the absorption spectra of 
finite PPA
and PDPA oligomers with eight double bonds along the polyene chain
for nonzero $H_{CB}$ ($t_{\perp}$ = 1.4 eV) but $H_{BB}$ = 0. Our choice
of finite oligomers is based on the finite conjugation lengths of the
existing materials \cite{Fujii}. 
The hopping integrals along the
polyene backbone here are 2.2 eV and 2.6 eV, corresponding to those of t-PA.
This is primarily because it is difficult to estimate the extent of the
bond alternations in finite chains, but also because
the polarizations and the oscillator strengths
of the various transitions depend weakly on the magnitudes of the hopping
integrals, and even the energies of the higher energy transitions do not 
depend  on these, as we have verified from explicit calculations.
We have indicated in Fig.~4 the polarizations of the different absorption
bands.
The
oscillator strength of the lowest energy chain $\to$ chain absorption is the
same in both cases, but that of the high energy molecular absorption 
(centered around 2$t_0$ = 4.8 eV) in PDPA
is nearly twice that in PPA. The large width of this absorption is due to
the nondegeneracy of the molecular transitions. At intermediate energy, the
weak second and third peaks correspond to chain HOMO -- 1 $\to$ LUMO + 1
and HOMO -- 2 $\to$ LUMO + 2 transitions, following which there occur the
charge-transfer transitions. 
The absorption
spectra for other $t_{\perp}$ can be guessed from 
Fig. 

\noindent\ref{huk-alphaxy-tb}(a) and (b).
Specifically, larger (smaller) the $t_{\perp}$, the broader (narrower) is the 
molecular absorption and the stronger (weaker) are the charge-transfer type
absorptions.

The absorption spectra for the case of nonzero
$H_{BB}$ are shown in Figs. \ref{huk-alphaxy-tb}(c) and (d). 
With increasing $t_{\sigma}$, the
molecular absorptions are broader, the charge-transfer transitions 
are  redshifted and acquire greater width.

\section{Optical Absorption for nonzero $H_{\lowercase{ee}}$}
\label{opt-sci}
There are two goals of our work in this section. First, we demonstrate that
the polarization characters of the lowest absorption in PPA and PDPA change
radically for nonzero $H_{ee}$, and that this change in the polarization
character provides the hint to understanding the mechanism of the
2A$_g$ -- 1B$_u$ crossover that
occurs in PDPA. Second, a strong $x$-polarized charge-transfer type absorption
appears at intermediate energy when both $H_{ee}$ and $t_{\sigma}$ are
nonzero. This intermediate energy absorption may have already been seen
experimentally in unoriented samples (see Fig.~2 in reference 
\onlinecite{Tada2}). 
In the following we discuss the energies and the
polarizations of the lowest energy absorption and the overall absorption spectra
separately.

\subsection{Lowest energy absorption.}

Our calculations of the optical transitions are within the 
SCI approximation. The SCI is known to give reasonable 
descriptions of the odd parity B$_u$ states (especially the lowest $B_u$
state, the 1B$_u$), even though it is a poor approximation for the even parity
A$_g$ states. In Table I we have given the optical gaps of PPA and PDPA
oligomers with eight double bonds for $H_{CB} \neq 0$, but $H_{BB}$ = 0
(reference \onlinecite{Shukla} discusses the case of five doubel bonds). 
These calculations are within the rigid bond approximation, with the hopping
integrals along the chain 2.2 eV and 2.6 eV, for single and double bonds,
respectively.
We show later that as within
one-electron theory, the role played by $H_{BB}$ in the low energy region is
weak. 

\begin{table}
\protect\caption{Optical gaps of oligomers of PDPA and PPA, 
containing eight double bonds in the backbone chain, within the H\"uckel
and the Coulomb correlated models, as a function of $t_{\perp}$. 
Calculations based upon the correlated model were performed within the 
SCI approximation. Note that the 
case of $t_{\perp} = 0.0$ corresponds to the eight double-bond polyene. All 
energies are in the units of eV.}
  \begin{tabular}{ccccc} 
 \multicolumn{1}{c}{$t_{\perp}$}  & \multicolumn{2}{c}{PPA} & 
\multicolumn{2}{c}{PDPA} \\
       & H\"uckel & SCI & H\"uckel & SCI \\ \hline
 0.00  & 1.45  & 3.26  &  1.45   & 3.26   \\
 0.35  & 1.44  & 3.17  &  1.42   & 3.08   \\
 0.70  & 1.40  & 3.13  &  1.35   & 3.01     \\
 1.40  & 1.28  & 3.02  &  1.14   & 2.80   \\
 1.80  & 1.20  & 2.94  &  1.01   & 2.65  \\
    \end{tabular}                      
  \label{opt-gap}    
\end{table}  

As seen in Table I, the optical gap decreases with $t_{\perp}$. Interestingly,
the decrease (relative to $t_{\perp}$ = 0)
is greater for nonzero $H_{ee}$ than for
$H_{ee}$ = 0, a result that is a consequence of increased delocalization
of the excited electron and hole of the 1B$_u$ into the phenyl 
components for nonzero $H_{ee}$, as we show below. 
The existing PPA and PDPA materials are finite
oligomers \cite{Fujii}, but for true long chain materials correct calculations
of the optical absorption should include self-consistent 
calculations of the
bond alternation (instead of the rigid bond approximation adopted here). In 
such a case, even larger decrease in the optical gap would be obtained.
In the existing literature it is often assumed that PPA and PDPA have larger
optical gaps than t-PA, but this is a consequence of the finite sizes of
the substituted materials. The optical gaps of the substituted materials
are smaller than the polyenes with the same chain length, with the gap in
PDPA smaller than that in PPA. 

In Table II we have given the $x$- and $y$-components of the transition
dipole couplings $\mu_x$  and $\mu_y$ 
between the ground state and the 1B$_u$ state of the
PDPA oligomer (the electronic charge is taken as 1, and the dipole couplings
are in units of \AA). The transition dipole coupling is known to be a direct
measure of the electron-hole correlation length in the 1B$_u$.
We do not show similar results for PPA, as the difference is
only quantitative. There are three interesting features of these
results. First, for each $t_{\perp}$,
$\mu_x$ decreases for nonzero $H_{ee}$, a result that is common to all
other $\pi$-conjugated polymers, and is a signature of exciton formation
and confinement of the electron-hole pair \cite{exciton1,exciton2}
in the 1B$_u$ in the $x$-direction
for nonzero $H_{ee}$. 
Second, $t_{\perp}$ decreases $\mu_x$ for both $H_{ee}$ = 0 and $\neq$ 0,
but this decrease is considerably larger for nonzero $H_{ee}$. For example,
the decrease in $\mu_x$ between $t_{\perp}$ = 0 and $t_{\perp}$ = 1.8 eV
is 2.6 \% for $H_{ee}$ = 0 but 8.2 \% for $H_{ee} \neq$ 0. This effect
is reminescent of the effect of $t_{\perp}$ on the optical gap (see above),
and is a signature of greater longitudinal electron-hole confinement for 
nonzero $t_{\perp}$ when $H_{ee}$ is nonzero. Note that with the Ohno Coulomb
parameters, which are intermediate in strength, relatively small increase in the
confinement can lead to 2A$_g$ -- 1B$_u$ crossover.  
In the case of linear chain polyenes within
the rigid bond approximation ({\it i.e.}, hopping integrals $t_0(1 \pm \delta)$)
the difference in the electron-hole correlation length in the 1B$_u$ state
\cite{Chandross2}.
between $\delta$ = 0.07 (E(2A$_g$) $<$ E(1B$_u$)) and $\delta$ = 0.3
(E(2A$_g$) $>$ E(1B$_u$)) is only 7 \%. In the present case, the increased
longitudinal confinement is accompanied, as well as driven, by the electron
correlation induced delocalization in the transverse direction, as observed
from the very interesting behavior of $\mu_y$. As seen in Table II
nonzero $H_{ee}$ {\it increases} $\mu_y$ 
for arbitrary {\it fixed} $t_{\perp}$, which is completely contradictory to 
the behavior in the $x$-direction. To the best of our knowledge, increasing
transition dipole coupling with $H_{ee}$ has not been found in any other
$\pi$-conjugated system. As noted in reference \onlinecite{Shukla} the
$\mu_y$ corresponding to an {\it isolated} trans-stilbene molecule oriented in
the transverse direction as in Fig.~1(b) decreases with $H_{ee}$, thereby
exhibiting the more common behavior. 
The opposite and surprising behavior in PDPA is a signature
of greater electron-hole 
delocalization in the $y$-direction for the correlated 1B$_u$ than for the
uncorrelated 1B$_u$. Taken together, these results indicate a synergistic
relationship between the confinement in the $x$-direction and the enhanced
delocalization in the $y$-direction:~ the delocalization in the $y$-direction
increases the confinement in the $x$-direction and vice versa.

The larger electron-hole delocalization in the $y$-direction is an electron
correlation effect that can be understood from the natures of the one-electron
levels. As shown in Fig. \ref{fig-chain-bond}(a), 
the contribution of the backbone chain atoms
to the one-electron wavefunction of the highest occupied one-electron level
(corresponding to the HOMO of the oligomer in the present context) decreases
with $t_{\perp}$. We have performed similar calculations for all one-electron
levels, and for all $t_{\perp}$, the contribution of the chain atoms to the
one-electron wavefunction decreases with increasing energy separation from
the chemical potential. Now, for $H_{ee}$ = 0, the 1B$_u$ is simply
the HOMO $\to$ LUMO transition. For $H_{ee} \neq 0$, the 1B$_u$ is a 
superposition of several transitions, including the HOMO $\to$ LUMO,
HOMO -- 1 $\to$ LUMO + 1, HOMO -- 2 $\to$ LUMO + 2, etc. The $y$-polarizations
of these higher energy excited configurations are larger
simply because the wavefunctions
themselves have smaller contributions from the chain atoms and larger
contributions from the phenyl group atoms, and as consequence the correlated
1B$_u$ wavefunction itself has a larger $y$-character. In Reference 
\onlinecite{Shukla}, we had
given a configuration space argument for this electron correlation induced
electron-hole delocalization in the $y$-direction. The present
k-space argument is essentially the same explanation, except that the
origin of the enhanced delocalization can be understood more 
quantitatively within these
k-space arguments.

As we have already pointed out in section III, increased confinement in the
longitudinal direction can in principle raise the energy of the 2A$_g$ with 
respect to the
1B$_u$. Although this still needs direct verification, the important
point here is that the increased confinement in the $x$-direction is a
consequence of the increased transverse
delocalization, which in turn {\it reduces} the optical gap (see Table I). 

\begin{table}
\protect\caption{The $x$ and $y$ components of $1A_g \protect\rightarrow 1B_u$ 
transition dipole moments $\mu_x$ and $\mu_y$ (in \AA), as a 
function of  $t_{\perp}$,
for a PDPA oligomer with eight double bonds. Note $t_{\perp}=0.0$ eV
corresponds to an eight double-bond polyene. }
  \begin{tabular}{lcccc} 
 \multicolumn{1}{l}{$t_{\perp}$}  &
 \multicolumn{2}{c}{$H_{ee}=0$} & 
\multicolumn{2}{c}{$H_{ee} \neq 0$ (SCI)} \\
       & $\mu_x$  & $\mu_y$ &  $\mu_x$   & $\mu_y$  \\
 0.00  & 3.07  & 0.35  &  2.08   &    0.73   \\
 0.35  & 3.06  & 0.40  &  1.97   &    0.92  \\
 0.70  & 3.04  & 0.54  &  1.96   &    0.99   \\
 1.40  & 3.00  & 0.97  &  1.93   &    1.21  \\
 1.80  & 2.99  & 1.24  &  1.91   &    1.34  \\
    \end{tabular}                      
  \label{edm-pdpa}    
\end{table}  

\begin{figure}[htb]
\twocolumn[\hsize\textwidth\columnwidth\hsize\csname @twocolumnfalse\endcsname
\epsfxsize=6.truein
\epsfysize=3.5truein
\centerline{\epsffile{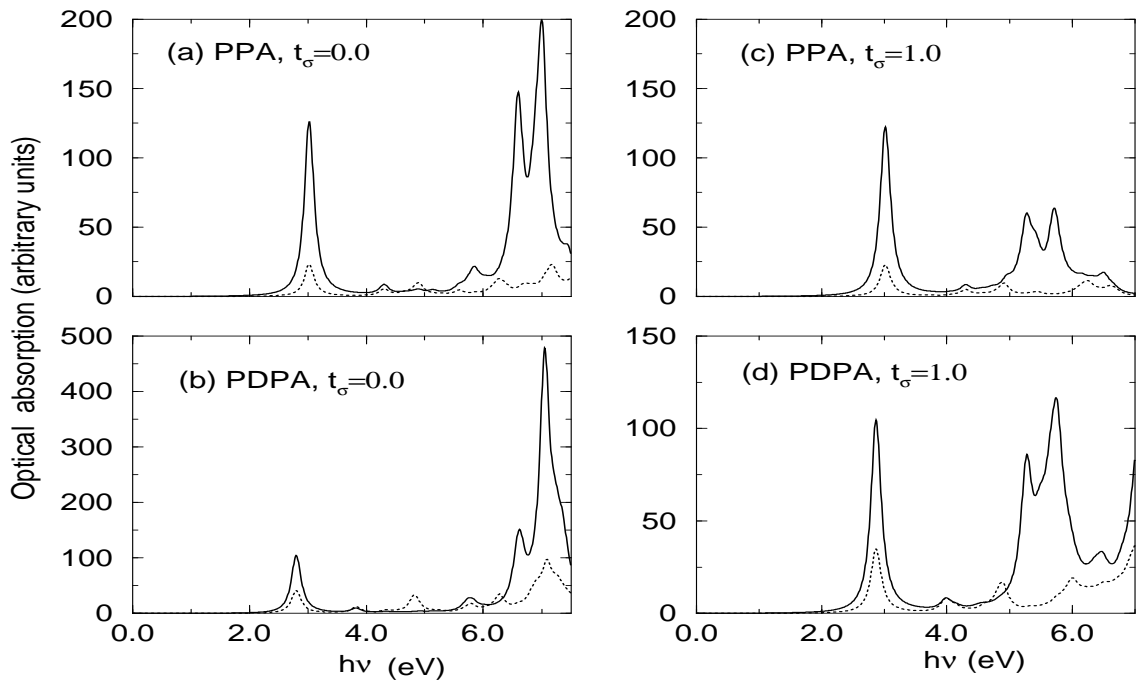}} 
\medskip
\caption{The calculated $x$ (solid line) and $y$ (dotted lines) components
of the optical absorption spectra
within the Coulomb correlated model for finite PPA (Figs.~(a) and (c)) and PDPA
(Figs.~(b) and (d)) oligomers with eight double bonds on the backbone polyene
chain. Note the different y-axis scales.
In all cases $t_{\perp}$ = 1.4 eV, while $t_{\sigma}$ = 0 in Figs.
(a) and (b) and 1.0 eV in Figs.~(c) and (d). Coulomb interactions increase the
$y$-character of the lowest absorption and the $x$-character of the
higher energy absorption bands.}
\label{sci-alphaxy-tb}]
\end{figure}

\subsection{Overall absorption spectrum}

In Figs. \ref{sci-alphaxy-tb}(a) and (b) 
we have shown the absorption spectra for PPA and PDPA
for the case of nonzero $H_{CB}$ but zero $H_{BB}$ for the specific case of
$t_{\perp}$ = 1.4 eV. We have included both $x$- and $y$-polarizations of the
absorption bands in these figures. With nonzero $H_{ee}$ there occur 
absorption bands at energies as high as 10 eV. These are not included in the
Figures. From comparison of these results with the absorption spectra
of Figs. \ref{huk-alphaxy-tb}
(a) and (b), the lowest absorptions in both PPA and PDPA have 
much larger contributions from the $y$-component here than for $H_{ee}$ = 0,
as has already been pointed
out in the previous subsection.

There is an even stronger difference between $H_{ee} = 0$ and $H_{ee} \neq 0$
in the high energy region. For $H_{ee} = 0$ the dominant absorption in the
high energy region is predominantly $y$-polarized, although there is also
substantial contribution from $x$-polarized transitions. In contrast, for
$H_{ee} \neq 0$, 
the strong absorption bands centered at $\sim$ 7 eV in both PPA and PDPA
in Figs. 5(a) and (b)
are not molecular absorptions but primarily charge-transfer chain $\to$
benzene and benzene $\to$ chain absorptions which have shifted to high
energies because of e-e interaction. The one-electron
chain-derived bands
are predominantly delocalized along the $x$-direction (with, however, 
substantial but smaller $y$-delocalization) and are also strongly hybridized
with the benzene $d$, $d^*$ MOs. The high energy absorption bands here 
originate primarily from transitions between these delocalized levels and
the benzene $l$, $l^*$ MOs, and as discussed in the context of one-electron
theory, these transitions are $x$-polarized. 

The charge-transfer type absorptions are redshifted further upon inclusion
of $H_{BB}$, as seen in Figs. \ref{sci-alphaxy-tb}(c) and (d), where we have
shown the optical absorptions for PPA and PDPA, respectively for $t_{\sigma}$
 =1.0 eV
(note that the $y$-contribution to
the oscillator strength of the lowest energy absorption to the 1B$_u$
has remained unchanged,
as was claimed in the previous subsection). The redshift of the charge-transfer
absorption also occurs within one-electron theory. However, within one-electron
theory, there occurs a strong overlap between the $x$-polarized charge-transfer
absorption and the $y$-polarized molecular absorption (see Fig. 
\ref{huk-alphaxy-tb})(c) and (d)).
In contrast, in the energy region of interest, there occurs only strongly
$x$-polarized charge-transfer absorptions, as the molecular absorptions are
now strongly blueshifted by e-e interactions. The surprising effect of the
e-e interactions is then that the absorption in the relatively high energy
region is more strongly $x$-polarized than the lowest energy absorption, even
though the latter is predominantly a backbone chain absorption.

Experimentally, a strong high energy absorption in a PDPA derivative has 
already been seen by Tada {\it et al.}, \cite{Tada2} but this is in an 
unoriented sample and the
polarization characters are unknown. We predict that in oriented samples this
absorption will be overwhelmingly $x$-polarized, which, as seen from 
comparisons
of Figs. \ref{huk-alphaxy-tb} and \ref{sci-alphaxy-tb}, is strictly 
a correlation effect. In the absence of
e-e interaction, the second band should have mixed $x$ and
$y$-character.

\section{$2A_{\lowercase{g}}$---$1B_{\lowercase{u}}$ ordering}
\label{2ag-1bu}

In our earlier Letter, we presented preliminary calculations on the relative
ordering of $2A_g$ and $1B_u$ excited states in PDPAs, using 
selected subsets of MOs of a two double-bond oligomer, within the full-CI (FCI)
approach~\cite{Shukla}.
Here we present more sophisticated calculations using
all the MOs of oligomers of PDPAs and PPAs with
four as well as six backbone carbon atoms, by employing the MRSDCI approach 
\cite{Tavan}. 
The MRSDCI is an efficient size-consistent algorithm that has
been shown to give energies of the 1B$_u$ and the 2A$_g$ that are as accurate
as the quadruple-CI (QCI) approach for polyenes up to sixteen carbon atoms
\cite{Tavan}, and has also been used to determine the frequency
dependent nonlinear optical properties of short polyenes \cite{Beljonne}.
Note that the PDPA oligomer with six backbone carbon atoms contains 42 carbon
atoms altogether, and the Hamiltonian matrices are much larger than those
previously investigated. As shown below, strong emphasis was placed on
obtaining very high accuracy, and the oligomer with six backbone carbons is the
largest that could be investigated without loss of accuracy.

Our calculations are within the rigid bond approximation, i.e., the hopping
integrals along the chain are taken to be $t_0(1 \pm \delta)$.
We know that enhanced bond alternation can lead to  $2A_g$ state being
higher than the $1B_u$ state \cite{Soos1,Soos2}. Therefore, in order to
understand the influences of electron correlations and
bond alternation on the $2A_g$-$1B_u$ ordering separately, we
considered backbone geometries of PPAs and PDPAs both with standard
polyene bond alternation
($\delta = 0.07$), 
and without any bond
alternation 
($\delta = 0.0$). 
For the case of PDPA oligomers,
full use of the $C_i$ symmetry of the system was made during the
correlated calculations. Therefore, $A_g$ and $B_u$ subspaces were
diagonalized in separate calculations \cite{ci-group}.  However, for the PPA
oligomers, 
because of the lack of any symmetry, $A_g$ and $B_u$ states were
obtained by diagonalizing the same Hamiltonian matrix.
The $1B_u$ state was identified in this case as the first
excited state with large dipole coupling with the ground state.
In the MRSDCI calculations presented below, we have ignored H$_{BB}$, based
on our results in Figs.~3, 4 and 5 
which clearly indicate that $H_{BB}$ plays an
insignificant role in the low energy physics. As far as the phenyl-chain
interaction term $H_{CB}$ is concerned, in all the calculations reported
below it was parametrized with $t_{\perp}=1.4$ eV.

The methodology behind the MRSDCI calculations is as follows.
The calculations are initiated with a Hartree-Fock (HF) computation of the
ground state of the oligomer concerned, followed by
a transformation of the Hamiltonian from the site representation to the
HF MO representation. Subsequently, a singles-doubles CI (SDCI) calculation
is performed, the three states of interest, viz., the $1A_g$, the $2A_g$, and
the $1B_u$ are examined, and the N$_{ref}$
configuration state functions (CSFs) making
significant contributions to their many-particle wave functions are
identified.  The next step
is the MRSDCI calculation for which the reference space consists of the
N$_{ref}$ CSFs identified in the previous step,
and the overall Hamiltonian matrix now includes configurations that are doubly
excited with respect to these reference CSFs (thereby including the
dominant quadruply excited configurations).
The new $1A_g$, $2A_g$, and
$1B_u$ states are now re-examined to identify new CSFs contributing 
significantly
to them so as to augment the reference space for the next set of MRSDCI
calculations. This procedure is repeated until
satisfactory convergence in the excitation energies of the $2A_g$ and
$1B_u$ energies is achieved. 
By the time convergence is achieved
typically all CSFs with coefficients of magnitude
$0.01$ or more in the corresponding many-particle wave functions have been
included in the MRSDCI reference space. Naturally, this leads to very large
CI matrices, e.g., the dimension of the CI matrix in the MRSDCI
calculation on the $B_u$ space of the three unit PDPA oligomer with
$N_{ref}=12$, was $\approx$ 2.22 millions. The lowest few eigenvalue and
eigenvectors of such large matrices were obtained by the Davidson
procedure~\cite{davidson}. The very large N$_{ref}$ required for convergence
(even for the 1B$_u$) is direct evidence of the contributions by the benzene
MOs to the correlated wavefunctions of PPA and PDPA.

The convergence pattern of our MRSDCI calculations with respect to N$_{ref}$
is demonstrated for the case of the PPA oligomer with
four backbone carbon atoms 
in Table \ref{tab-ppa2} \cite{note3}. 
These calculations correspond to $\delta$ = 0.
The $2A_g$ and $1B_u$ excitation energies
of various intermediate MRSDCI calculations are measured with respect
to the best $1A_g$ ground state obtained in the largest MRSDCI calculation
the table ($N_{ref}=25$). It is for this reason
that with  increasing  $N_{ref}$, the $2A_g$ and $1B_u$ excitation
energies decrease monotonically. It is clear from the behavior of
excitation energies with respect to N$_{ref}$
depicted in Table \ref{tab-ppa2} that by the time the MRSDCI calculation
with the largest value of N$_{ref}$ is performed, the 2A$_g$ and 1B$_u$
excitation energies  have converged to an acceptable accuracy.

Our overall MRSDCI results for 2A$_g$ and 1B$_u$ energies for four and six
backbone atom PPA and PDPA oligomers (both with and without bond alternation),
along with the corresponding final N$_{ref}$'s, are summarized in
Table \ref{tab-all}. In the same Table, for the sake of comparison, the
results of full CI calculations on four and six atom polyenes, both with and
without bond alternation, are also presented.
Upon examining the data presented in Table \ref{tab-all}, the following
trends emerge: (i) in PDPA oligomers without backbone bond alternation,
$1B_u$ and $2A_g$ states are nearly degenerate, while, in the same oligomers
with bond alternation, the $1B_u$ state is significantly below the $2A_g$, (ii)
in PPA oligomers without bond alternation the $2A_g$ state is below the
$1B_u$, while for oligomer geometries with bond alternation
$2A_g$ is very slightly above the 1B$_u$, (iii) 
in all cases, the dominant effect of the phenyl-substitution is the lowering
of E(1B$_u$), which decreases as t-PA $>$ PPA $>$ PDPA. This last result is in
agreement with the SCI results for much longer oligomers in 
Table \ref{opt-gap}, confirming our statement that the SCI captures the
essential nature of the 1B$_u$ state.

The occurrence of the 2A$_g$ slightly below the 1B$_u$ in the 
PDPA oligomer with $\delta$ = 0 and six backbone carbons is 
not surprising. As discussed in reference \onlinecite{Shukla}, exactly as
bond-substituted systems can be thought of as effective linear chains
with large bond alternation, site-substituted materials can be thought of
also as linear chains, but now with composite sites (as opposed to bonds) with
small effective on-site Coulomb repulsion $U_{eff}$. In linear chains of
sufficient length, E(2A$_g$) $<$ E(1B$_u$) for $\delta$ = 0 for any $U_{eff}$.
The more important point, however, is that the energy gap 
E(1B$_u$) -- E(2A$_g$) has decreased from 1.15 eV to 0.02 eV in going from
the polyene with $\delta$ = 0 to the PDPA oligomer. As discussed in section
IV.A, the persistence of the Fermi surface degeneracy for $\delta$ = 0,
even upon 
phenyl-substitution (see Figs.~2(a) and (b)), ensures that unconditional 
bond alternation occurs in both PPA and PDPA, and therefore the nonzero 
$\delta$ results in Table \ref{tab-all} are more appropriate for the real
systems.
Although our calculations are for relatively short
oligomers, we expect our conclusions regarding
the $2A_g$---$1B_u$ ordering in PDPAs to be valid
even for long chains (in particular, for the experimental systems, which are
also finite, albeit of length that is twice the lengths studied here). 
To begin with, the size-dependence of the 2A$_g$ -- 1B$_u$ energy gap, for
the case E(2A$_g$) $>$ E(1B$_u$) is steepest at the smallest sizes (a result
easily confirmed within one-electron theory). Nevertheless, 
E(2A$_g$) -- E(1B$_u$) has almost converged already for the PDPA oligomers
with $\delta \neq$ 0: for the oligomer with four backbone carbon atoms this
energy difference is 0.4 eV, while for the oligomer with six backbone carbons
the difference is 0.37 eV. Second, our calculations are for rather large
Coulomb interactions and fairly small $t_{\perp}$, while in the real systems
the Coulomb interactions are perhaps smaller \cite{Chandross1} and $t_{\perp}$
might be larger. Finally, it is clear from Table \ref{tab-all} that even if
the 2A$_g$ were to be below the 1B$_u$ in short PDPA oligomers, the energy
difference would be tiny. Density matrix renormalization group calculations
for linear chain systems have suggested that specifically in these cases
there is a size-dependent 2A$_g$ -- 1B$_u$ crossover, where in longer chains
E(2A$_g$) $>$ E(1B$_u$) again \cite{Shuai,Barford}. Thus from all possible
considerations, the lowest two-photon state is above the lowest one-photon
state in PDPA. In contrast to the PDPA oligomers, E(2A$_g$) -- E(1B$_u$) has
not converged in the PPA oligomers, where this number is 0.13 eV for the
oligomer with four backbone carbons and 0.06 eV in the oligomer with six
backbone carbons. It is conceivable therefore that in slightly longer 
oligomers, the 2A$_g$ is slightly below the 1B$_u$ in PPA.

\section{Concluding discussions}
\label{conclusion}

We begin this section with the summary of results obtained in this paper.
We agree with the authors of the experimental
reference \onlinecite{Hidayat} that light
emission in PDPAs involves predominantly the backbone chain and not the
trans-stilbene moiety. However, as we have shown here, the lowest absorption
has a strong $y$-character, which is enhanced by e-e interactions, and it
is this enhanced delocalization in the $y$-direction that is ultimately the
origin of the 2A$_g$ occurring above the 1B$_u$ in these systems. 
\begin{table}  
\protect\caption{Convergence pattern of MRSDCI $2A_g$ and $1B_u$ energies for
PPA with four backbone carbon atoms and uniform hopping integrals
 with respect to increasing $N_{ref}$.
$N_{ref}$ denotes the total number
of reference configurations included in the MRSDCI calculation in
question.  The $2A_g$ and
$1B_u$ energies are with respect to the $1A_g$ ground state obtained
in the largest calculation ($N_{ref}=25$). All energies are in eVs.}
\protect
  \begin{tabular}{cccc}
System    &    $N_{ref}$ &  $E(2A_g)$   &   $E(1B_u)$   \\ \hline
PPA &     2  & 5.50 & 4.77 \\
 & 6              &    4.94  &            4.69 \\
 & 8              &    4.74 &           4.65 \\
& 10          &        4.67  &           4.65 \\
& 12          &        4.60   &           4.63 \\
& 15          &        4.52 &            4.62 \\
& 19         &         4.50 &            4.61 \\
& 21        &          4.46 &            4.60 \\
& 23       &           4.43 &            4.60 \\
& 25       &           4.42 &            4.59 \\
   \end{tabular}
  \label{tab-ppa2}    
\end{table}

\begin{table}  
\protect\caption{The best MRSDCI results for $2A_g$ and $1B_u$ energies of
the oligomers of PDPA and PPA containing four  and six
carbon atoms  along the polyene backbone. 
Backbone configurations
with both $\delta =
0.0$ and $\delta = 0.07$, were considered.
For the sake of comparison, the results of the full CI calculations (indicated
by FCI in the $N_{ref}$ column) for
the two states of the corresponding polyenes are also given. Since
for PPA oligomers the $A_g$ and $B_u$ states were obtained in the same
calculations,  the same values of $N_{ref}$ are indicated
for both the $A_g$ and $B_u$ states.}
\protect
  \begin{tabular}{cccccccc}
        &       & \multicolumn{6}{c}{Energy Gaps (eV)} \\
        &       & \multicolumn{3}{c}{Four backbone carbons} &
\multicolumn{3}{c}{Six backbone carbons} \\
 System & State & $N_{ref}$ & $\delta = 0.0$ & $\delta = 0.07$ &
$N_{ref}$ & $\delta = 0.0$ & $\delta = 0.07$ \\ \hline
PDPA    & $2A_g$ & 17 & 4.38 &  4.98 & 13 & 3.62 & 4.39  \\
        & $1B_u$ & 8  & 4.25 &  4.58 & 12 & 3.64 & 4.02 \\
PPA     & $2A_g$ & 25 & 4.42 &  5.03 & 14 & 3.66 & 4.47 \\
        & $1B_u$ & 25 & 4.59 &  4.90 & 14 & 4.02 & 4.41 \\
polyene & $2A_g$ & FCI & 4.66 &  5.35 & FCI &3.51 & 4.37 \\
        & $1B_u$ & FCI & 5.49 &  5.80 & FCI & 4.66 & 5.02 \\
   \end{tabular}                      
  \label{tab-all}    
\end{table} 

\noindent The
authors of reference \onlinecite{Hidayat} emphasize exciton confinement due
to relaxations involving phonons. As shown here, enhanced exciton confinement
along the backbone is a consequence of the electron-hole delocalization in the
transverse direction. This does not imply that electron-phonon interactions
are unimportant. Rather, even if relaxations involving 
phonons play a strong
role in the experimental systems, the 2A$_g$ had to be already close 
to the 1B$_u$ in the PDPAs (given the large
separation between these in t-PA) due to a purely electronic mechanism, and
the present work gives the details of this electronic mechanism.
We emphasize that the longitudinal
confinement does not necessarily imply enhanced exciton binding energy, as the
delocalization in the transverse direction will also reduce the energy
of the conduction band threshold. The bond alternation along the 
backbone decreases with phenyl substitution, also 
as a direct consequence of smaller
chain atom contributions to the one-electron levels at and near the Fermi
level. Inclusion of the overlaps between the benzene MOs that occur due
to phenyl ring torsions does not change the results of any  
calculation that probes the low energy photophysics.
Finally, we predict a strong second high
energy absorption that will be predominantly $x$-polarized in oriented
materials, in contradiction to the prediction of one-electron theory. 
Experimental work can therefore directly verify the role of e-e interactions.

In the case of PPAs, our results are more difficult to interpret. The
convergence behavior of E(2A$_g$) and E(1B$_u$) in the two oligomers that
have been studied suggests that here E(2A$_g$) $<$ E(1B$_u$) in longer
chains, although the energy difference can be very small.
The small energy difference between these levels would imply that even 
small perturbations can either reverse this ordering, or lead to strong
state mixing. We speculate that the behavior of the PAPAs is related to this
proximity between the 2A$_g$ and the 1B$_u$. PAPAs with small alkyl groups
probably are similar to the PPAs of the same length and are therefore weak
emitters. Large alkyl groups either lead to greater main chain bending
(and therefore greater exciton confinement) and state reordering, 
or because of their larger
electron donating capability they cause a greater electronic perturbation
leading to greater state mixing between the 2A$_g$ and the 1B$_u$.

In our investigation of optical properties here, we have focused on ground
state absorption only, and theoretical studies of excited state absorptions
\cite{Gontia,Epstein} is a topic of future research. In the case of PPP and
PPV, it has been claimed that the occurrence of multiple classes
of one-electron bands leads to
several different photoinduced absorptions in  ultrafast
pump-probe spectroscopy \cite{Chakrabarti}, and it is conceivable that similar
results will be found here too.

We make one final remark regarding the future application of this work for
the design of novel materials. Photo- and electroluminescence of
$\pi$-conjugated polymers are currently the focus of intense research, and
research on the design of novel light emitting diodes using these materials
has already reached a mature stage \cite{Friend}. Initial work has also
shown promising results for laser applications using these materials
\cite{Friend1,Hide,Valy}. In all these cases, however, the materials being
used correspond to those obtained by bond substitution, and the optical gaps
of such systems are necessarily larger than that of t-PA. Our demonstration
that light emission can be a consequence of enhanced delocalization in a 
direction transverse to the direction of extended conjugation, and that the
optical gaps in this latter class of systems are smaller than linear polyenes 
of the same
length, introduces the interesting possibility of designing novel rigid long
chain light emitting $\pi$--conjugated polymers with optical gaps smaller
than that in t-PA. The actual synthesis of rigid materials may be difficult,
as the finite conjugation lengths of PPAs and PDPAs already demonstrate, but 
nevertheless, the present work shows this to be conceptually feasible.
Infrared dye lasers are extremely rare, and the possibility
of obtaining new light sources at telecommunications wavelengths and in the
eye-safe spectral region should make this an exciting new research direction.

\acknowledgements
Work in Arizona was supported by NSF-ECS and the ONR through the MURI center
(CAMP) at the University of Arizona. We acknowledge valuable discussions
with Z.V. Vardeny and M.J. Rice.

%
%
%
%
%

\begin{references}
\bibitem{note1} The phenyl groups themselves may have additional substituted
side groups, like n-butyl etc. Thus there exist many different PDPAs, see
reference \onlinecite{Tada2}. However, the behavior of all these systems
are nearly identical, indicating that it is only the conjugation with the
bare phenyl groups that modifies the behavior of the phenyl-substituted
polyacetylenes.
\bibitem{ci-group} The $C_i$ group has two irreducible representations:~
$A_g$ and $A_u$. The two-photon states in this group belong to the
$A_g$ representation, while the one-photon states belong to $A_u$
representation. However, in the present
paper we refer to the one-photon states of PDPAs as the $B_u$ states in order
to facilitate direct comparison to the similar states in polyenes.
\bibitem{note3} In this small molecule the 1B$_u$ and the 2A$_g$ energies are
rather high and therefore close to the $d \to l^*$ and $l \to d^*$ excitations
of the phenyl segments. However, from transition dipole couplings and the
natures of the configurations that form the many-body wavefunctions the
identifications of the 1B$_u$ and the 2A$_g$ are simple and unambiguous.
\end{references}
\end{document}